\documentstyle[11pt]{article}
\catcode`\@=11

\@addtoreset{equation}{section}

\catcode`\@=11
     
\def\section{\@startsection{section}{1}{\z@}{3.5ex plus 1ex minus
   .2ex}{2.3ex plus .2ex}{\large\bf}}

\newcommand{\be}{\begin{equation}}
\newcommand{\ee}{\end{equation}}
\newcommand{\bea}{\begin{eqnarray}}
\newcommand{\eea}{\end{eqnarray}}

\newcommand{\half}{{1\over2}}
\def\Idoubled#1{{\rm I\kern-.22em #1}}
\newcommand{\IR}{\Idoubled R}
\newcommand{\DA}{D_{(A)}}
\newcommand{\req}[1]{Eq.\ (\ref{#1})}

\newcommand{\spa}{\Sigma_{n-1}}
\newlength{\savebls}
\begin{document}
\setlength{\savebls}{\baselineskip}
\setlength{\baselineskip}{1 \savebls}
\begin{titlepage}
\begin{flushright}
UNB Technical Report 99-02\\
WATPHYS-TH99/01
February 1,1999\\
\end{flushright}
\vspace{.25in}

\begin{center}
{\large\bf Gravity From Topological Field Theory}\\
\vspace{15pt}
{\it by}\\
\vspace{5pt}
J.\ Gegenberg \\[5pt]
{\it Department of Mathematics and Statistics}\\
   {\it University of New Brunswick}\\
   {\it Fredericton, New Brunswick
   \it Canada E3B 5A3}\\
{[e-mail:  lenin@math.unb.ca]}\\
\vspace{13pt}
R.\ B.\ Mann\\ [5pt]
{\it Department of Physics}\\
{\it University of Waterloo}\\
{\it Waterloo, Ontario Canada N2L 3G1}\\
{\it and}\\
{\it Institute for Theoretical Physics}\\
{\it University of California Santa Barbara}\\
{\it Santa Barbara, California}\\
{\it USA 93106-4030}\\
{[e-mail:mann@itp.ucsb.edu]}\\
\end{center}
\vspace{20pt}
{\bf Abstract}\\
We construct a topological field theory 
which, on the one hand, generalizes BF theories in that there is 
non-trivial coupling to 
`topological matter fields'; and, on the other, generalizes the 
three-dimensional model of Carlip and Gegenberg to arbitrary 
dimensional manifolds.   Like the three dimensional model, the theory can 
be considered to describe a gravitational field interacting with topological 
matter.  In particular, in two dimensions, the model is that of gravity on a 
torus.  In four dimensions, the model is shown to admit  
constant curvature black hole solutions.
\end{titlepage}

\section{Introduction}

The considerable amount of progress made in recent years in quantum gravity
has in large part resulted from advances in $(2+1)$-dimensional general 
relativity.
In the absence of matter, the field equations imply that spacetime is
locally flat.  Consequently the theory has no propagating modes, and the
reduced phase space is finite dimensional, allowing one to fully 
quantize the theory \cite{Carlip}.  Unfortunately,
the addition of matter has the general effect of destroying this 
exact solvability, introducing propagating modes which render the
phase space infinite dimensional.  

However there is a theory in which a class of matter fields in $(2+1)$ dimensions
does not destroy solvability.  Referred to as BCEA theory,
it minimally couples a pair of topological matter fields 
(one-forms $B^a$ and $C^a$) to gravity in such a way that
the connection $A^a$ remains locally flat \cite{cargeg}, while
allowing the triad $E^a$ to have non-trivial structure.  Hence, 
gravitational and matter degrees of freedom are inextricably mixed.  In
addition, the field equations admit constant
curvature black hole solutions that are analogous (but not
identical) to the $(2+1)$-dimensional BTZ black hole \cite{bceabh}.
In the context of BCEA theory, the role of inner and outer event 
horizons for the (spinning) BTZ black hole are interchanged \cite{fat}. 

Furthermore BCEA theory has a supersymmetric generalization for which
one can show that this black hole solution is 
supersymmetric \cite{papa}.  The 
supersymmetrized theory can be quantized, and 
it turns out that the partition function is 
a topological invariant which in a sense contains both 
the sum of the squares of 
the Ray-Singer analytic torsions and the Casson invariant 
of flat SO(3) bundles over the three 
dimensional spacetime manifold \cite{cargeg,geg}.

In this paper we construct higher-dimensional versions of BCEA
theory, which we call BQPA theories.  The class of BQPA theories
minimally couple topological matter (i.e. matter which does not
couple directly to a metric) to gravity.  The metric is a derived 
quantity which can have non-trivial structure due to the presence
of the matter fields.  In particular, black holes of
constant negative curvature are solutions to the field equations,
with the (negative) cosmological constant being a constant of
integration.  These black holes are analogous (but not
identical) to the class of topological black holes recently
considered in the literature \cite{amin,robbtop,banados}.

The outline of the paper is as follows.  In Section 2, we briefly 
review the original 3D BCEA theory.  In Sections 3 and 4 we develop 
the higher dimensional generalization- BQPA theory.  In Section 5 we 
show how BQPA theory may be considered as a theory of gravity interacting 
with (topological) matter fields.  In Section 6 we explicate the slightly 
distinct case of the two dimensional theory, showing, in particular, that 
it can be considered as a theory of gravity on a torus, with the various 
topological matter fields related to the complex geometry of the torus.  
In Section 7  the four dimensional theory is shown 
to admit cosmological solutions and in Section 8 
constant curvature black hole solutions are obtained.  We close with some 
some speculations and suggestions for further work.

\section{BCEA Gravity}

The BQPA theory is the generalization to n dimensions of a 3-D topological 
field theory 
\cite{cargeg}, 
which we have called ``BCEA theory'' \cite{bceabh}.  
In this section, we review BCEA theory, highlighting those features useful 
in the generalization to  spacetimes with arbitrary dimension.

The action for BCEA theory is
\be
S[B,C,E,A]=
-\half\int_{M^3}\left(E^i\wedge F_i[A]+B^i\wedge \DA C_i\right) ,
\label{bceaaction} 
\ee
where $\DA$ is the covariant derivative with respect to the SO(3) (or 
SO(2,1)) 
connection $A^i$ with curvature $R_a[A]$ given by
\be
F_i[A]:=dA_i+\half[A,A];\label{curv}
\ee
and $E^i,B^i,C^i$ are 1-form fields over $M_3$ taking their values in the 
Lie algebra so(3), respectively so(2,1).  In particular, $A=A^i G_i$, with 
$G_i$ the generators of the Lie group:
\be
[G_i,G_j]=\half\epsilon_{ijk}G^k,\label{so3alg}
\ee
where the indices $i,j,...=0,1,2$ are raised and lowered with respect to 
the Kronecker delta $\delta_{ij}$ in the case of SO(3), and the Minkowski 
metric $\eta_{ij}=\hbox{diagonal}(-1,+1,+1)$ in the case of SO(2,1).

The stationary points of $S[B,C,E,A]$ are determined by the field equations
\bea 
F_i[A]&=0, \nonumber\\
\DA B^i&=0, \nonumber\\
\DA C_i&=0, \nonumber\\
\DA E^i+\half \epsilon^{ijk}B_j\wedge C_k&=0.
\label{3deqm} 
\eea
It is because of the term in $B_j\wedge C_k$ in the 
last equation of motion, that the
triad $E^i$ is not, in general, compatible with the locally flat
spin-connection $A_i$.
Nevertheless, the equations of motion above determine a 
non-trivial spacetime geometry 
on $M^3$: if we define a one-form field $Q_i$ by the requirement
\be
\epsilon^{ijk}\left(Q_j\wedge E_k -B_j\wedge C_k\right)=0 ,
\label{aa0}
\ee
then the equation of motion for the $E^i$ can be written as
\be
dE^i+\half\epsilon^{ijk}\omega_j\wedge E_k=0,\label{tor} 
\ee
where
\be
\omega_i:=A_i+Q_i .
\label{omega}
\ee
\req{tor} may be recognized as the condition that the frame-field $E^i$
be compatible with the (nonflat) spin connection $\omega_i$.

We may thus interpret BCEA theory as a model of (2+1)-dimensional gravity
with a triad $E^i$ and a connection $\omega_i$ coupled to matter fields
$B^i$ and $C^i$.  
The geometry is determined by the
metric $g_{\mu\nu} = \eta_{ij}E^i{}_\mu E^j{}_\nu$ in the case where the 
gauge group is SO(2,1), and with $\eta_{ij}$ replaced by $\delta_{ij}$ 
for SO(3).

The action functional $S[B,C,E,A]$ is invariant under a twelve-parameter group
whose infinitesimal generators are \cite{cargeg}
\bea
\delta B^i&=&\DA \rho^i+\half\epsilon^{ijk}B_j\tau_k,\nonumber\\
\delta C^i&=&\DA\lambda^i+\half\epsilon^{ijk}C_j\tau_k,\nonumber\\
\delta E^i&=&\DA\xi^i+\half\epsilon^{ijk}\left(E_j\tau_k
 +B_j\lambda_k+C_j\rho_k\right),\nonumber\\
\delta A^i&=&\DA\tau^i .
\label{3dsym} 
\eea
This group may be recognized as I(ISO(2,1)), where the notation IG denotes 
the semi-direct product of the Lie group G with its own Lie algebra 
${\cal L}_G$.  Like the action for ordinary Einstein gravity in three
dimensions \cite{witten}, the BCEA action can be obtained from a
Chern-Simons functional, now for the gauge group I(ISO(2,1)).

Although BCEA theory has not been widely explored, a few of its
properties are known.  Quantization of the theory has been discussed:
one can show \cite{cargeg} that the partition function is  
given by the sum of the squares of 
the Ray-Singer torsions for each flat connection $A$.  Black hole
solutions
of various types were later obtained  as solutions of the equations 
of motion \req{3deqm} \cite{bceabh}, including an analogue of the BTZ black 
hole \cite{btz}.  Although the metric formally is the same as that
of the BTZ metric,  in BCEA theory the constants of integration
have a different interpretation, and
the black hole thermodynamics  is not of the Bekenstein-Hawking form,
except in the extremal case \cite{bceabh}. 

\section{BQPA Theory}

We now generalize the BCEA theory to $n$ dimensions.  
The group involved is the Poincare group 
$I'G$ built from the $n$-dimensional 
Lorentz group, $G= SO(n-1,1)$ (or the rotation group $G=SO(n)$).  If the 
generators of $G$ are $G_{ij}$, with $i,j,...=0,1,...,(n-1)$ and $T_j$ 
generate translations, then the set $\{G_{ij},T_k\}$ generates $I'G$:
\bea
\left[G_{ij},T_k\right]&=&\half\left(g_{jk}\delta_i^l-g_{ik}\delta^l_j\right)
T_l;\nonumber\\
\left[G_{ij},G_{kl}\right]&=&\half\left(g_{kj}G_{il}+g_{jl}G_{ki}+g_{ik}G_{lj}
+g_{il}G_{jk}\right);\nonumber
\\
\left[T_i,T_j\right]&=&0,\label{i'g}
\eea
where $g_{ij}:=\eta_{ij}$ for $SO(n-1,1)$ and $g_{ij}=\delta_{ij}$ for $SO(n)$.
The simplest way to 
do this is via the action functional
\be
S[B,Q,P,A]=-\half\int_{M_n}P^{ij}\wedge F_{ij}[A]+\DA B^i\wedge Q_i.
\label{ndaction}
\ee
Here, $A:=A^{ij}G_{ij}$ is an SO(n-1,1) (or SO(n)) connection.  The curvature  
$F^{ij}[A]=\half F^{ij}_{\mu\nu}[A]dx^\mu dx^\nu$ is given by 
\be
F^{ij}[A]=dA^{ij}+A^i{}_k\wedge A^{kj}.\label{ndcurv}
\ee
The 1-form fields $B=B^i T_i$ and $(n-2)-$form fields $Q=Q^i T_i$ take their 
values in the
translation subgroup of $I'SO(n-1,1)$ while the $(n-2)-$form fields 
$P=P_{ij}G^{ij}$ take 
their values in the dual to the Lie algebra generated by the $G_{ij}$.

The exterior covariant derivatives with respect to the $SO(n-1,1)$ connection $A$ 
of the
`translation' and `rotation' type fields are, respectively,
\bea
\DA B^i&=&dB^i+A^i{}_j\wedge B^j,\nonumber\\
\DA P_{ij}&=&dP_{ij}+A_i{}^k\wedge P_{kj}+A_j{}^k\wedge P_{ik}.
\eea

The stationary configurations are given by the classical equations of motion
\bea
F^{ij}[A]&=&0,\label{F}\\
\DA P_{ij}+B_{[j}\wedge Q _{i]}&=&0,\label{P}\\
\DA B^i&=&0,\label{B}\\
\DA Q_i&=&0.\label{Q}
\eea

{}From the analysis \cite{bithesis} it is straightforward to show that the 
symmetry group of
the theory is $I(I'SO(n-1,1))$, i.e., the semi-direct product of $I'SO(n-1,1)$ 
with its Lie algebra.  Indeed, the infinitesimal gauge transformations which 
preserve the
action \req{ndaction} are given by \cite{bithesis}
\bea
\delta A^{ij}&=&\DA\tau^{ij},\label{deltaA}\\
\delta Q_i&=&\DA\xi_i-\tau_i{}^j Q_j,\label{deltaQ}\\
\delta B^i&=&\DA\lambda^i-\tau^i{}_j B^j,\label{deltaB}\\
\delta P_{ij}&=&\DA\nu_{ij}+\lambda_{[i}Q_{j]}-\xi_{[i}B_{j]}+2\tau_{[i}{}^k
P_{j]k}.\label{deltaP}
\eea
The gauge parameters are form fields, with $\tau^{ij}=-\tau^{ji}$ and 
$\lambda^i$ both 0-form
fields, while $\xi^i$ and $\nu_{ij}=-\nu_{ji}$ are $(n-3)-$form fields.  For $n>3$ 
the gauge
symmetry is reducible.

We emphasize here that the action \req{ndaction} is a topological field theory 
--  like Chern-Simons or BF theories there are no propogating modes -- in 
other words the (reduced)
phase space is a finite dimensional manifold.  Also, as in other topological field 
theories,
the diffeomorphisms in the `base manifold' $M_n$ are gauge transformations if the
equations of motion are satisfied \cite{bithesis}.

\section{Hamiltonian Analysis}\label{sec4}

In the 3-D theory, the 1-form fields $E^i$ are the momenta canonically conjugate 
to the
connection components $A_i$ \cite{cargeg}.  In the Ashtekar formulation of 
canonical
general relativity theory, the spatial triads are conjugate to the complex 
connections
\cite{ash}.  This motivated extracting gravitational physics from the 3-D 
theory by
identifying the $E^i$ with the spacetime triad.  In the following we generalize 
the
Hamiltonian analysis of the 3-D theory to $n$-dimensions as a prelude to 
extracting
gravitational physics.  We generalize the results obtained by Bi \cite{bithesis} 
for $n=4$.

Locally the splitting of the manifold $M_n$ into space 
$\Sigma_{n-1}$
and time $\IR$ is accomplished by defining a volume element $\epsilon^{a_1... 
a_{n-1}}$
on $\Sigma_{n-1}$ from the permutation symbol $\epsilon^{\mu_1...\mu_n}$ on $M_n$ 
by 
\be
\epsilon^{a_1...a_{n-1}}=\epsilon^{0 a_1...a_{n-1}}.\label{vol}
\ee
Let $t$ denote the `time coordinate' on $\IR$ and $x^a$ the `spatial coordinates' 
on a
patch on $\Sigma_{n-1}$.  Then after integrating terms in $\partial_0 A_{ija}$ and
$\partial_0 B^i_a$ by parts and using the notation $\dot f:=\partial_0 f, 
D_af^i:=\partial_a
f^i + A^i{}_j f^j$ we get for the action \req{ndaction}
\bea
S&=&{1\over2 (n-2)!}\int dt\int_{\spa} d^{n-1}x
\left\{
\epsilon^{a_1...a_{n-2}a}\left[\dot
A_{ija}P^{ij}_{a_1...a_{n-2}}+\dot B^i_a Q_{i a_1...a_{n-2}}+\right.
\right. \nonumber\\
&+&\left.\left. A_{ij0}
\left(D_aP^{ij}_{a_1...a_{n-2}}
+B^j_aQ^i_{a_1...a_{n-2}}\right)+
B^i_0D_aQ_{ia_1...a_{n-2}}\right]+\right.\nonumber\\
&+&\left.(n-2)!\epsilon^{a_1...a_{n-3} a b}
\left[\half P^{ij}_{0a_1...a_{n-3}}
F_{ijab}+Q_{i0a_1...a_{n-3} }D_aB^i_b\right]
\right\}
+\nonumber\\
&+&{1\over2(n-2)!}\int dt\oint_{\partial\spa}dS_a\epsilon^{a_1...a_{n-2} 
a}\left(A_{ij0}P^{ij}_{a_1...a_{n-2}}
+B^i_0Q_{i a_1...a_{n-2}}\right).\label{splitaction}
\eea
The first two terms give the momenta canonically conjugate to $A_{ija}$ and 
$B^i_a$ 
respectively
\bea
\Pi^{ija}&:=&{\delta S\over\delta\dot A_{ija}}={-1\over 
(n-2)!}\epsilon^{a_1...a_{n-2}
a}P^{ij}_{a_1...a_{n-2}},\\
\Pi^a_i&:=&{-1\over2(n-2)!}\epsilon^{a_1...a_{n-2} a}Q_{ia_1...a_{n-2}}.
\eea

On the other hand, the coefficients of the Lagrange multipliers
$A_{ij0}, B^i_0$, $P^{ij}_{0a_1...a_{n-3}}, Q_{iaa_1...a_{n-3}}$ 
are the constraints
\bea
C^{ij}&:=&D_a\Pi^{ija}+B^{[j}_a\Pi^{i]a}\approx 0,\label{cij}\\
C_i&:=&D_a\Pi^a_i\approx 0,\label{ci}\\
C_{ijab}&:=&F_{ijab}\approx 0,\label{cijab}\\
C^i_{ab}&:=&D_{[a}B^i_{b]}\approx 0.\label{ciab}
\eea
It can be shown that the Poisson bracket algebra satisfied by the constraints is 
the Lie
algebra of the group $I(I'SO(n-1,1))$.  See \cite{bithesis} for the case $n=4$.

Now the $\Pi^{ija}$ are Lie algebra valued components of a spatial vector field 
density. 
The most natural way to extract a {\it spatial} frame-field is 
\be
\tilde E^{Ia}={(-1)^n\over[(n-2)!]^2}\epsilon^{a_1...a_{n-2}a}P^{0I}_{a_1...
a_{n-2}}
.\label{tildeE}
\ee
The indices $I,J,...=1,2,..., (n-1)$ denote components in the Lie algebra 
$so(n-1)$ of the
little group of $SO(n-1,1)$.  The $\tilde E^{Ia}$ are the densitized spatial 
components of
the contravariant frame-fields.  Note that $\tilde E^{Ia}$ is the spatial 
Hodge dual of the spatial components of the 
$(n-2)-$form $P^{0I}$.  We will use this in the next section 
to construct Lorentzian 
(or Riemannian) spacetime geometries.  We note here that, 
only in the case $n=4$, is it 
equally natural to define the densitized contravariant spatial triad by
\bea
*\tilde E^{Ia}&:=&\half \epsilon^I{}_{JK}\Pi^{JKa}\nonumber\\
&=&-{1\over2}\epsilon^I{}_{JK}\epsilon^{abc}P^{JK}_{bc}.\label{*tildeE}
\eea

\section{The Gravitational Sector of BQPA Theory}\label{sec5}

We saw in Section 1 that the 3-D theory could be interpreted as a gravity plus 
gauge field
theory by identifying the 1-form fields $E^i,\omega^i$ with the spacetime geometry 
for any
configuration in which the $\omega^i$ are solutions of the algebraic equations 
\req{aa0}. 
We shall generalize this to the $n-$dimensional case in this section.

In the case $n>3$, the the field $P^{ij}$, corresponding to $E^i$, is an 
$(n-2)-$form field
and, in general, we saw that a {\it densitized} spatial contravariant frame-field 
$\tilde
E^{Ia}$  can be defined via \req{tildeE}.  Now we define 
in the case that $n\geq 3$ a contravariant vector frame-field
$E^{Ia}$ by
\be
E^{Ia}:=\tilde E^{-1/n-2}\tilde E^{Ia},\label{e}
\ee
in a region where $\tilde E\neq 0$, with $\tilde E:=\det(\tilde E^{Ia})$.
\footnote{For the case of $n=2$, this construction fails.  See the 
following section for the case $n=2$.}   In this region,
$E^{Ia}$ has an inverse which we denote by $E^I_a$.  It follows that
$E:=\det(E^I_a)=1/\det(E^{Ia})$, so that $E=\tilde E^{1/n-2}$.

A Riemannian metric $h_{ab}$ on $\spa$ is then given by the the inverse of 
\be
h^{ab}=\delta^{IJ}E^a_I E^b_J.
\ee
The spacetime geometry is specified, in the ADM formulation, by the spatial 
frame-field $E^I=E^I_a dx^a$ 
together with the lapse function $N$ and shift vector $N^a$ on $\spa$.  
\be
ds^2=-N^2dt^2+(\delta_{IJ}E^I_a E^J_b)(dx^a+N^adt)(dx^b+N^bdt)
\ee
We define the lapse and shift in terms of the $P^{ij}$ via
\bea
N&=&{1\over (n-1)[(n-2)!]^2}{\epsilon^{aba_1...a_{n-3}}\over E}
E_{Ia}e_{Jb}P^{IJ}_
{0a_1...a_{n-3}}
,\label{N}\\
N^a&=&{(-1)\over[(n-2)!]^2}{\epsilon^{aba_1...a_{n-3}}\over E}E_{Ib}
P^{0I}_{0a_1...a_{n-3}}
.\label{Na}
\eea

We have defined the ADM components- the spatial metric, lapse function 
and shift 
vector- from some of the components of the form $P^{ij}$.  We can contruct 
a {\it covariant} frame-field $e_\mu^i$ from the ADM components via 
\be
e^I_a=E^I_a,\quad e^I_0=E^I_a N^a,\quad e^0_0=N.
\ee
Note that the components $e^0_a$ are not determined from the ADM 
variables.  One may use the gauge freedom to choose, say, $e^0_a=0$.
This is consistent with
\footnote{We use the convention that if $w$ is a p-form, then $w={1\over p!}
w_{a_1...a_p}dx^{a_1}\wedge...\wedge dx^{a_p}$.}
\be
P^{ij}=\epsilon^{ij}{}_{k_1...k_{n-2}}e^{k_1}\wedge...\wedge e^{k_{n-2}},
\label{Pfrome}
\ee
as long as the frame-field satisfies $e^0_a=0$, and then it follows 
that 
$P^{IJ}_{a_1...a_{n-2}}= 0$.   
Note that the $N,N^a$ are Lagrange multipliers enforcing constraints in both 
cases.

In the case $n=4$, the components $P^{IJ}_{ab}$ are subject to an 
intriguing interpretation. 
 As we saw at the end
of Section 3- \req{*tildeE}- we could define the densitized spatial frame field 
with respect
to the components $P^{IJ}_{ab}$ rather than the $P^{0I}_{ab}$.  Now 
on-shell the $A_a^{ij}$- 
the spatial
components of the $SO(3,1)$ connection- are {\it flat}.
One can choose a gauge in which the $A_a^{ij}$ are self-(or anti-self-)dual, i.e. 
\be
*A_a^{ij}:=\half\epsilon^{ij}{}_{kl}A_a^{kl}=\pm A_a^{ij}.\label{selfdual}
\ee
The corresponding components of the canonical momentum $\Pi^{ija}$ are 
self-(anti-
self-)dual.  Hence the densitized spatial frame fields satisfy 
\be
*\tilde E^{Ia}=\pm\tilde E^{Ia}.
\ee
Hence, up to a gauge transformation, the two geometries defined by 
$*\tilde E$ and $\tilde E$ are equivalent.  This  contruction 
is reminiscent of the formulation of 
Einstein gravity in terms of the Ashtekar variables \cite{ash}.

\section{The Two Dimensional Case}

The two dimensional case is completely and explicitely solvable, though, as 
we will see, somewhat degenerate.  It is informative to display the solution 
for this case, since it illustrates many of salient features of the BQPA 
theories.  

We will show below that, in the simple case where the spacetime is
two dimensional, that the gauge group of the BQPA model is $I'SO(2)$.
We note here that this case is rather singular compared with that of
higher spacetime dimension, where the gauge group is $I(I'SO(N))$. 

In this case, $P_{ij}=\epsilon_{ij}P$ is a 0-form.  The 
indices $i,j,...=0,1$.    
The $SO(2)$-connection is $A_{ij}=\epsilon_{ij}A$.  The $Q_i$ and $B^i$ 
are respectively 0-and 1-form fields taking values in so(2).  

Since the group $SO(2)$ is abelian, the vanishing curvature condition gives, 
in a coordinate patch $U$, 
\be
A=d\alpha,
\ee
where $\alpha$ is a 0-form field.  Now define 
\bea
Q&:=&Q_0+iQ_1,\nonumber\\
B&:=&B_0+iB_1.
\eea
Then the equations of motion for $Q_i$ and $B^i$ can be written respectively 
in forms
\bea
dQ&=&id\alpha Q,\nonumber\\
dB&=&id\alpha B,
\eea
and have respective general solutions
\bea
Q&=&q e^{i\alpha},\nonumber\\
B&=&(\bar B+d\beta) e^{i\alpha},
\eea
where $q:=q_0+iq_1$ is an arbitrary complex constant, $\bar B^i$ is a harmonic 
1-form field and $\beta^i$ is an arbitrary 0-form field, with $\bar B=\bar B^0
+i\bar B^1$ and similarly $\beta=\beta^0+i\beta^1$.

The equations of motion for $P^{ij}$ reduce to
\be
dP+\half(Q_0B^1-Q_1B^0)=0.
\label{dP}\ee

It is straightforward to show that in the Hamiltonian analysis there are two 
first class constraints:
\bea
C&:=&2\partial_x P+\epsilon^i{}_jB^j_xQ_i,\\
C_i&:=&\partial_xQ_i+\epsilon_i{}^jA_xQ_j.
\eea
It follows that the Poisson bracket algebra is precicely that of the 
Lie algebra of the two dimensional Poincare group $I'SO(2)$.

We now proceed to construct a spacetime geometry from the 
$I'SO(2)$ structure.  We are of course dealing with the Riemannian case.  
We restrict to the case that the 2-manifold is closed and compact.  
Indeed, 
it is not surprising that our $I'SO(2)$ structure forces the torus topology 
onto the 2-manifold, 
since the isometry group of the torus is precicely $I'SO(2)$.  The form of 
\req{dP} suggests that $dP$ is the imaginary part of the product $\half \bar Q
B$, where $\bar Q$ is the complex conjugate of $Q$.  Hence we define
\bea
d\hat P&:=&-\half \bar Q B\\
&=&-\half\left(Q_iB^i+\epsilon^i{}_jQ_iB^j\right).
\eea  
We see that $d\hat P$ is closed if $B$ and $Q$ are closed under
the exterior covariant derivative $D_A$ with respect to the flat
Abelian connection $A$.  Furthermore, the explicit dependence on the flat
connection $A$ in $B$ and $Q$ cancel in $\bar Q B$.  We construct $B$ 
locally from the 
Beltrami form $dz+\mu d\bar z$ and the flat connection $A=d\alpha$ as 
follows: 
\be
B=2e^{i\alpha}\left(dz+\mu d\bar z\right). 
\label{belB}\ee
Note that $B$ above is a non-trivial harmonic 1-form field if $\partial_z\mu
=0$.  Then 
\be
d\hat P=\bar q\left(dz+\mu d\bar z\right).\label{belP}
\ee
This gives the usual form for the metric on 2-torus:
\be
ds^2=\mid d\hat P\mid^2=\mid q\mid^2\mid dz+\mu d\bar z\mid^2.\label{belM}
\ee
With $\mu=$constant, the metric is that of a flat torus with Teichm\"uller  
parameter $\tau$ determined by $\mu=(1+i\tau)/(1-i\tau)$. 
Hence, $\hat P$ is the complex coordinate in which the metric 
on the torus is isothermal.

\section{Cosmological Solutions}

We consider in this section spacetime solutions to the $BQPA$ theory.  These are
straightforward to obtain. For a spacetime of constant positive curvature the
solution corresponding to the metric
\be\label{desit}
ds^2 = -dt^2 + e^{2t/l}\left(dx^2+dy^2+dz^2\right),
\ee
is given by
\bea
P_{ij} &=&\epsilon^{ijkl} e_k\wedge e_l, \label{PP}\\
B_I = 0  &\quad&  B_0 =- 4 e^{2t/\ell}dt, \\
Q_I =  {1\over\ell}\epsilon_{IJK}dx^J\wedge dx^K  &\quad&  Q_0 = 0, \\
A_i{}^j& =& 0,\label{AA}
\eea
where 
\be
e^0 = dt \qquad  e^I = e^{t/l} \delta^I_adx^a,
\ee
and $I,J,...=1,2,3$.  

Likewise the solution with constant negative curvature 
\be\label{adesit}
ds^2 = dz^2 + e^{2z/\ell}\left(-dt^2+dx^2+dx^2\right),
\ee
is given by \req{PP} and \req{AA}, but with  
\bea\label{desitsol}
B_I = 0  &\quad&  B_3 = 4 e^{2z/\ell} dz ,\\
Q_I =  {1\over\ell}\epsilon_{IJK}dx^J\wedge dx^K  &\quad&  Q_3 = 0 , \\
\eea
where 
\be\label{tetrads}
e^I = e^{z/\ell} \delta^I_a dx^a \qquad    e_3 = dz ,
\ee
with indices $I,J,...=0,1,2$ raised and lowered by the $2+1$-dimensional 
Minkowski metric $\eta^{IJ}=diagonal(-1,+1,+1)$.   

The solutions derived above illustrate an application of our approach toward
deriving gravity from topological field theory
in the simplest non-trivial case, namely that of a spacetime with
constant curvature.

\section{Black Holes}

The method for obtaining black hole solutions from spacetimes of constant negative
curvature are by now well-established \cite{btz,amin,robbtop}. Since spacetimes of 
constant negative curvature are solutions to the field equations,
we can expect that black hole spacetimes are solutions to the field equations as 
well. 
We find that this is indeed the case. We obtain the solution
\begin{eqnarray}\label{bhP}
&\left[P_{ij}\right] 
=& \nonumber\\
&
\left[ 
{\begin{array}{cccc}
0\,& \,{\displaystyle -2R^2\,\sinh{\theta}}\,{\it d\theta\wedge d\phi}\,& 
\,{\displaystyle \frac {2R}{\sqrt{f}}\,\sinh{\theta}}
\,{\it dR\wedge d\phi}\,
& 
\,{\displaystyle \frac {-2R}{
\sqrt{f}}}\,{\it dR\wedge d\theta} 
\\ [2ex]
 2R^2\sinh{\theta}\, {\it d\theta\wedge d\phi}\,&\,0\, &\,{\displaystyle 
 2\,R\,\sqrt{f}\sinh{\theta}}\,{\it dT\wedge d\phi}
 \,& \,
{\displaystyle -2\,R\,\sqrt{f}}\,{\it dT\wedge d\theta}
\\ [2ex]
 {\displaystyle \frac {-2R}{\sqrt{f}}}\,{\it dR\wedge d\phi}
\, & \, {\displaystyle -2R\sqrt{f}\,\sinh{\theta}}\,{\it dT\wedge d\phi}
 \, & \,0\, & \,2{\it dT\wedge dR}\,
\\ [2ex]
\,{\displaystyle \frac {2R}{\sqrt{f}}}\,
{\it dR\wedge d\theta}\, & {\displaystyle 2R\,\sqrt{f}}
\,{\it dT\wedge d\theta} 
\, &  -2\,{\it dT\wedge  dR}\, & 0
\end{array}}
\right]\nonumber\\
 &&\end{eqnarray}
\be\label{bhB}
\left[B_i\right] = 
-4\,e^{2T/\ell}\sqrt{f}\left({\it dT}+\frac{R}{\ell\,f}{\it dR}\right) \left[
\,1\,,\,-\frac {R}{\ell}\,,\,0\,,\,0\,\right],
\ee
and
\be\label{bhM}
\left[Q_i\right] ={\displaystyle  \frac {2\,R\,}{\ell\,f}\,\,e^{\frac{-2T}{l}}}
\left[{\displaystyle \frac{R^2}{\ell\,\sqrt{f}}\,\sinh{\theta}} {\it d\theta\wedge d\phi}
,{\displaystyle  -\frac{R}{\sqrt{f}}\,\sinh{\theta}}\,
{\it d\theta \wedge d\phi} 
 , {\displaystyle \sinh{\theta}}\,\lambda
\wedge {\it d\phi}, 
{\displaystyle -\lambda
\wedge {\it d\theta} }
\right].
\ee 
In the above, the quantity $f:=-1+R^2/\ell^2$ and 
$$
\lambda:= 
\left(\frac {R}{\ell}\,{\it dT}+\frac {1}{f}\,
{\it dR}\right).
$$

The connection is 
\be\label{bhgauge} 
A_i{}^j :=  \left[
{\begin{array}{cccc} 0 &  - {\displaystyle \frac {{\it dR}}{\ell\,f}}
& -{\displaystyle \frac {R{d\theta}}{\ell\,\sqrt{f}}}  & - {\displaystyle
\frac {R\,\sinh{\theta}{d\phi} }{\ell\,\sqrt{f}}}  \\ [2ex]
 - {\displaystyle \frac {{\it dR}}{\ell\,f}}  & 0 & {\displaystyle \frac
{d\theta}{\sqrt{f}}}  & {\displaystyle \frac 
{\sinh{\theta}}{\sqrt{f}}\,{\it d\phi}}  \\ [2ex] 
-{\displaystyle \frac
{R}{\ell\,\sqrt{f}}}{\it d\theta}  & -{\displaystyle \frac {{\it
d\theta}}{\sqrt{f}}}  & 0 & -\cosh{\theta} {d\phi}
\\ [2ex]
 - {\displaystyle \frac {R}{\ell\,\sqrt{f}}\,\sinh{\theta}}\, {\it d\phi}
&  - {\displaystyle \frac {l}{\sqrt{f}}\,\sinh{\theta}}\, {\it d\phi}
&  \cosh{\theta}\, {\it d\phi}\, & 0 \end{array}}
 \right] \ee and it is straightforward to show that it is flat, i.e.
$F^{ij}[A]=dA^{ij}+A^i{}_k\wedge A^{kj} = 0$.

Using the prescription given in
section (\ref{sec5}) we find that $P^{ij}=\epsilon^{ijkl} e_k\wedge e_l$, consistent with
eq. (\ref{Pfrome}), and so the metric is
\begin{equation}\label{e1}
    ds^2 ={\displaystyle  
    - f\,{\it dT}^2 + f^{-1}\,{\it dR}^2 
      + R^2\,\left({\it d\theta}^2 + \sinh^2(\theta)\, {\it d\phi}^2\right)},
\end{equation}
where $l = \sqrt{3 / |\Lambda|}$, with $\Lambda$ the (negative) cosmological 
constant.
The $(\theta,\phi)$ section of
the metric (\ref{e1}) describes a compact space of genus $g>2$
provided $\theta$ and $\phi$ are appropriately identified.  The section described
by these coordinates is that of a hyperbolic space or
pseudosphere. This space becomes compact upon identifying the opposite edges of 
a $4g$-sided polygon (whose 
sides are geodesics) centered at the origin $\theta=\phi=0$ of the pseudosphere.
An octagon is the simplest such polygon, yielding a compact space of genus
$g=2$ under these identifications. Further details of this construction
can be found in refs. \cite{amin,robbtop}.

Once these identifications are made, the metric (\ref{e1}) describes the spacetime
of a topological black hole of genus $g>2$, with event horizon at $r=l$.  Indeed,
the above solution (\ref{bhP}--\ref{bhgauge}) is a $(3+1)$ dimensional 
generalization of the 
black hole solution obtained in  BCEA theory \cite{bceabh}. Note that
this is a local solution which yields the metric (\ref{e1}).

\section{Summary}

We have in this paper demonstrated how one can extend to higher dimensions
the understanding of gravity derived from a topological field theory
that has been previously given in $(2+1)$ dimensions \cite{cargeg,bceabh}.
In these BQPA theories the connection is flat and 
the metric is a derived quantity. However the field equations
give rise to  spacetime structures that are (at least) as interesting
as in $(2+1)$ dimensions, as shown in the
previous two sections. We suggest that this latter feature -- in conjunction
with the finite number of degrees of freedom these models have --
make them attractive candidates to study, particularly in terms of
their quantum properties.

A number of interesting questions remain.  What is the physical interpretation
of the remaining degrees of freedom $P^{IJ}_{a_1...a_{n-2}}$ ?  In 
particular, are there topological obstructions to the `gauge' $P^{IJ}_{a_1...
a_{n-2}}=0$?  
Are there other interesting solutions to the theory which have
a clear physical interpretation?  How does the coupling of other
forms of matter affect the basic physical properties and structure
of BQPA theories?  What resemblance do these theories have to their
$(2+1)$ dimensional counterparts?  These and other topics remain the subjects
of future investigation.

\bigskip\noindent
{\bf ACKNOWLEDGEMENTS}

\noindent
We would like to acknowledge the partial support 
of the Natural Sciences and Engineering Research Council of Canada
and of the National Science Foundation (Grant No. PHY94-07194). 
R.B. Mann would like to thank the hospitality of the ITP in Santa
Barbara, where some of this work was carried out.

\end{document}